Scientific
Research
Publishing

# The remnant of neutron star-white dwarf merger and the repeating fast radio bursts

**Xiang Liu[1,2]**

[1] Xinjiang Astronomical Observatory, Chinese Academy of Sciences, 150 Science 1-Street, 830011 Urumqi, China

[2] Qiannan Normal University for Nationalities, Longshan Street, 558000 Duyun, China
  Email: liux@xao.ac.cn

## Abstract

Fast radio bursts (FRBs) at cosmological distances still hold concealed physical origins. Previously Liu (2018) proposes a scenario that the collision between a neutron star (NS) and a white dwarf (WD) can be one of the progenitors of non-repeating FRBs and notices that the repeating FRBs can also be explained if a magnetar formed after such NS-WD merger. In this paper, we investigate this channel of magnetar formation in more detail. We propose that the NS-WD post-merger, after cooling and angular momentum redistribution, may collapse to either a black hole or a new NS or even remains as a hybrid WDNS, depending on the total mass of the NS and WD. In particular, the newly formed NS can be a magnetar if the core of the WD collapsed into the NS while large quantities of degenerate electrons of the WD compressed to the outer layers of the new NS. A strong magnetic field can be formed by the electrons and positive charges with different angular velocities induced by the differential rotation of the newborn magnetar. Such a magnetar can power the repeating FRBs by the magnetic reconnections due to the crustal movements or starquakes.

## Keywords

## 1. Introduction

An increasing number of fast radio bursts (FRBs) have been discovered and cataloged [1]; for recent results see [2], but their origins remain a mystery ([3-4] for reviews). Their large dispersion measures (DMs) which typically exceed the Galactic DM (estimated by using the models of [5-6]) suggest an extragalactic origin. The millisecond duration of the FRBs implies their emission-region sizes must be smaller than those of the magnetosphere of neutron stars (NS). However, at a cosmological distance, the typical flux density of 1 Jy of observed FRBs corresponds to luminosity up to $10^{43}$ erg/s [7], which has never been observed in normal pulsars. Most of the FRBs are observed as one-time burst-like events. If such FRBs are intrinsically non-repeating, they should be produced by cataclysmic events when a supermassive neutron star collapses to form a black hole [8-9] or during the collisions among black hole (BH), NS and white dwarf (WD) [10-12].

In a total of about 100 FRBs, only two non-repeating FRBs (namely FRB 180924 and FRB 190523) have recently been localized at galaxies with very low star formation rate [13-14]. Ten FRBs have been observed to be repeating sources [15-17], which require a long-lived central engine. Only one of the repeating FRBs, namely FRB 121102, has been localized with a star forming host galaxy to at redshift of 0.193 ([18] and references therein). Detailed follow-up observations show that the source is highly magnetized [19], likely in an environment of a magnetar [20-22]. The repeating pulses of FRB 121102 show complex frequency and time structure [23]. A variety of models have been proposed for the repeating FRBs [24-25], some of which involve magnetars. There are 29 magnetars observed in the Milky Way and The Magellanic Clouds (See the McGill magnetar catalog[1][26]), though none of them shows high luminosity pulses as observed in FRB 121102, probably because of FRBs only present at the very early age of magnetars. It is also possible that there are different magnetars formed from different channels. However, the formation mechanisms of magnetars are not well investigated yet.

The magnetic field of magnetars is about $10^{14-15}$ G, which is about 100-1000 times greater than that of the ordinary pulsar formed in the supernova explosion. Magnetars are thought to be formed by the core-collapse of the highly magnetized massive stars [27-28] or by the NS-NS mergers [29-30] or probably by the accretion induced collapse (AIC) of WD and binary WD merger [31-32]. In our Galaxy, 11 out of 32 massive stars sample show magnetic fields of 1000-5000 G [33], which would be able to result in a strong magnetic field of the final collapsar product for the magnetic flux conservation. The first confirmed NS-NS merger was detected in a multi-messenger manner in the event GW170817 by the gravitational detectors LIGO/Virgo (GW), Fermi/GBM (EM), and followed by multi-wavelength ground telescopes. The remnant of the NS-NS merger in the GW170817 event remains an open question [34]. Previous works have estimated the magnetar formation rate at 1-10% of all pulsars or even 40% by assuming their short lifetime (magnetic decay time) of $\sim 10^4$ yr [35].

It is proposed that the collision of an NS and a WD can be one of the progenitors of non-repeating FRBs and claimed that repeating FRBs may be produced by a magnetar formed in the collapse of an NS-WD merger [12]. In this paper, we investigate this process of magnetar formation in more detail. Depending on the total mass of the two merging objects, the product of an NS-WD merger may form either (i) a black hole or (ii) a magnetar or (iii) a hybrid WDNS. For case (ii), as suggested by [24-25], it can power repeating FRBs by the magnetic reconnections due to the crustal movements or starquakes of the magnetar.

## 2. Three possible remnants of neutron star-white dwarf mergers

Simulations by [36-38] suggest that there are two fates for close NS-WD binaries. One of them is stable mass transfer (SMT) that the matter from the WD is accreted across the inner Lagrange point onto the NS, and the binary will evolve in a secular timescale for inspiral. The other is an unstable mass transfer (UMT) through tidal disruption, which will lead to a fast inspiral and collision of the NS into the WD in a hydrody-namical timescale [36-37]. In the second possibility, reference [37] suggested that the NS will plunge into the WD and spiral to its center if the white dwarf is massive enough, which may form a quasi-equilibrium configuration that resembles a Thorne-Zytkow object [39]. The simulations show that the NS-WD system may merge through the UMT process when the WD mass is greater than 0.2 $M_\odot$ [38]. The merger rate of the NS-WD binaries will be dominated by the UMT rate because the timescale of the UMT is significantly shorter than that of the SMT for the binary to merge [36-37].

---

[1]http://www.physics.mcgill.ca/~pulsar/magnetar/main.html

However, the evolution and final remnant of NS-WD merger have not been explored in detail. It has been suggested that a Thorne-Zytkow-like object (TZO) will form at first (note the different scales: we consider here an NS merged into a WD, not an NS merged with a red giant star), and if the total mass of the TZO > 2.5 $M_\odot$ it will finally collapse into a black hole [37]. If the total mass of the TZO is less than 2.5 $M_\odot$, a collapse is also possible to take place following cooling and angular momentum redistribution [37], which will ultimately form a new NS or a spinning quasi-equilibrium configuration of a cold NS surrounded by a hot mantle of the WD. Except the BH product, we analyze further the NS-WD post-merger products with $M < 2.5\,M_\odot$ in the following.

If the total mass $M < 2.5\,M_\odot$ but sufficiently greater than the Chandrasekhar limit of 1.44 $M_\odot$, a new NS may be formed via the collapse of the NS-WD merger after cooling down and angular momentum redistribution [37]. This can be considered as follows. For an NS (1.4 $M_\odot$) merged with a heavy WD (e.g. > 0.6 $M_\odot$), the electron degenerate force, thermal pressure force and centrifugal force cannot support the gravitational force. The WD will collapse onto the NS to form a new NS.

The origin of the strong magnetic field of NS is not well understood yet. For an NS formed from the core collapse of a massive star during a supernova explosion, the magnetic field of the NS have resulted from the core of progenitor star via the magnetic flux conservation. The stellar magnetic field can be formed, e.g. by a magneto hydrodynamic dynamo [40]. The magnetic field of the NS can be an inheritance of the stellar magnetic field. However, as noted by [41], the inheritance is not sufficient to explain the fields of NSs and magnetars. The magnetar was first proposed in [29] and analyzed by [42-43]. They suggested that an efficient helical dynamo action could produce the dipole magnetic field of $10^{14-15}$ G in the magnetars formed by core-collapse massive stars or by the merger of an NS binary.

In this paper, we propose a magnetar formation scenario from the collapse of an NS-WD post-merger. If the mass of the NS-WD merger is sufficiently large, after cooling down, the electron degenerate pressure, thermal and centrifugal forces cannot support the gravity. The WD will then collapse onto the NS and forms a new NS. The newly formed NS is effectively a magnetar, because a strong magnetic field could be formed as introduced below.

The magnetic fields of NS can be produced by the positive and negative charges inside the NS. Considering that the newly formed NS from the NS-WD merger rotates differentially, and the enormous degenerate electrons from the WD compressed into the outer layers of the new NS, so the electrons and positive charges (mainly protons) are distributed differently. Assuming that the electrons are mostly distributed in the outer layers of the new NS with total charges $Q_e$ and average angular-velocity $\omega_e$, and that the protons are mostly distributed in the main body of the NS with total charges $Q_p$ and average angular-velocity $\omega_p$. We can derive the $B$-field outside the NS in the spherical coordinate:

$$\mathbf{B} = \frac{(Q_p * \omega_p + Q_e * \omega_e)R_0^2}{3c}\left[\frac{2cos\theta}{R^3}\boldsymbol{e_r} + \frac{sin\theta}{R^3}\boldsymbol{e_\theta}\right]\,, \qquad (1)$$

where $R_0$ is the new NS radius, $R$ is the radial distance from the center of NS, $c$ is the speed of light.

Considering the neutrality of total charges in the NS, i.e. $Q_p = -Q_e$, to produce a $B$-field for an NS, the angular velocities of the electrons and protons $\omega_e$, $\omega_p$ should be different in equation (1). For instance, assuming that the electrons are in relativistic regime with an average `stream' speed of 0.1c, i.e. an angular velocity of $\omega_e \sim 3 \times 10^3$ rad/s, and that the protons are not relativistic with an angular velocity close to the NS spin (i.e. $\omega_p \sim 1$ rad/s for the spin of $2\pi$ in a few seconds period), we can estimate $|Q(\omega_p - \omega_e)| \sim 3 \times 10^3 Q$ in equation (1) with $Q_p = -Q_e = Q$. To produce $10^{11-12}$ G at the surface of NS for normal pulsars, the Q is calculated to be $\sim 1.5 \times 10^{20}$ C equivalent to $\sim 10^{39}$ electrons (about the amount of fully ionized electrons from 1.7 billion tons of atomic hydrogen $\sim 10^{-18} M_{\odot}$, a tiny mass for a WD or NS) for $\theta = 0$. If we take a smaller speed ($< 0.1$ c) of the electron stream, more electrons ($> 10^{39}$) are required to produce the magnetic field of $\sim 10^{11-12}$ G. Thus, for a certain angular-velocity difference above, in order to create $\sim 10^{14-15}$ G for magnetars, the charges Q has to be 100-1000 times larger than that in normal NS, which may be viable for the magnetar formed in the collapse of NS-WD merger, where enormous degenerate electrons from the WD are compressed into the new NS. In addition, the metallicity of the merger would also play a role in contribution of charges and conductivity when the matter was highly ionized in the merger product.

In principle, from equation (1) the sum of $Q_p\omega_p + Q_e\omega_e$ can be positive, zero, or negative, when $\omega_p$ is larger, equals to or less than $\omega_e$, respectively, for the neutrality of charges ($Q_p = -Q_e = Q$). So the dipole magnetic field of an NS can change in direction by up to $180°$ (reversal) or even disappears when the term $Q_p\omega_p + Q_e\omega_e$ equals to zero. The $B$-field reversal and disappearance might occur in old pulsars when their differential rotation becomes marginal and their positive and negative charges are more evenly distributed inside the NS with similar angular velocities. Some phenomena observed in old pulsars, e.g. ``nulling", may be caused by such an effect.

In addition, we discuss the spin of the magnetar formed by the collapse of an NS-WD merger. The angular momentum (*Jns,wd*) of an eccentric NS-WD binary is *Jns,wd=Jns+Jwd+Jorb*, which including the angular momentum of NS, WD, and the orbital term. And also *Jns,wd=Jns'+Jwd'+Jloss*, where *Jns'* and *Jwd'* is respectively the angular momentum of NS and WD after the binary merged. *Jloss* is the sum of angular momentum losses from ejecta/wind (*Jej*), gravitational wave (*Jgw*), electromagnetic waves (*Jem*), neutrino emission (*Jneu*), and that dissipated into internal thermal energy or heat (*Jheat*) in the merger:

$$Jloss=Jej+Jgw+Jem+Jneu+Jheat \tag{2}$$

When the post-merger product finally collapses to form a new NS or a magnetar, according to the angular momentum conservation, we have

$$Jns,wd=Jns'+Jwd'+Jloss=Jmag+Jloss+Jloss' \ , \tag{3}$$

where *Jmag* is the angular momentum of the magnetar, and *Jloss'=Jej'+Jgw'+Jem'+Jneu'+Jheat'*, the *Jloss'* in the final collapse contains similar components of angular momentum loss as in .equation (2). Angular momentum of the putative magnetar is then

$$Jmag=Jns'+Jwd'-Jloss'=Jns+Jwd+Jorb-Jloss-Jloss' \tag{4}$$

In combining all the processes above, the *Jns,wd* of the merger will be significantly lost from the ejecta (e.g. 14-18% of initial total mass escapes to infinity [37]), gravitational wave radiation, neutrino emission, etc.

In equation (4), when the $Jloss+Jloss'$ approaches to the $Jorb$, we expect that the $Jmag \sim Jns+Jwd \leq 2Jns$ (for $Jwd \leq Jns$), i.e. the spin of the magnetar would be comparable to that of the proto-NS before merging. If the $Jloss+Jloss' \ll Jorb$, the magnetar would spin much faster than the proto-NS.

In the newborn magnetar, the differential rotation energy ($E_{\Delta\Omega}$) can be converted to magnetic energy $E_B$ [41], but it is not very clear how this conversion operates. In our scenario in this paper, the $E_{\Delta\Omega}$ can naturally increase the difference of the angular velocities between the electrons and positive charges that could be distributed differently from the collapse of the NS-WD merger, thus can increase efficiently the magnetic field in equation (1) by the differential rotation (angular velocities) of electrons and positive charges in the magnetar.

If the total mass is not significantly larger than the Chandrasekhar limit, e.g. an NS merged with a smaller mass WD (in 0.2 - 0.6 $M_\odot$), after cooling and accretion of the WD debris [44], the merger product may finally form a spinning quasi-equilibrium configuration consisting of a cold NS and a hot mantle (a hybrid WDNS, [37]). It will not collapse to form a new NS since the mass of the WDNS is not large enough.

## 3. Discussion on emission and formation rate of remnants

As analyzed above, there are three kinds of remnants from the NS-WD merger: first, a stellar BH; second, a new NS or a magnetar; third, a hybrid WDNS. These remnants may exhibit some transient emissions during and after their formation, and we discuss these remnants further including the magnetar formation rates as follows.

1) When the NS-WD post-merger collapses to form a stellar BH, an outburst could be observed during the collapse. A disk and a plasma corona around the BH [45] may be formed from the extended WD debris and fallback of ejecta, and produce afterglow emission.

2) When the NS-WD post-merger collapses to form a new NS, multi-wavelength outbursts are expected. Some of the remnants may form magnetars as we analyzed. From the newborn magnetar, repeating FRBs can be created by the coherent curvature radiation of the electrons [46-47] energized through the magnetic reconnections by the crustal movement where the magnetic fields are anchored [24], or starquakes [25]. The crust of very young magnetar is unstable due to the differential rotation, significant magnetic decay and spin-down of the magnetar.

3) The NS-WD merger product will never collapse to form a BH or a new NS if the total binary mass is not large enough. Instead, it will finally form a hybrid WDNS after cooling and accreting of the WD debris. Multi-wavelength emissions are expected from the hybrid WDNS for the higher surface temperature and magnetic field excited by the NS core.

We discuss roughly the magnetar formation rate from the NS-WD merger. The time-integrated rate of NS-WD mergers is 3-15% of the type Ia supernovae rate [48]. The type Ia SN rate is $\sim 8 \times 10^4$ $Gpc^{-3}yr^{-1}$ in z < 0.5 [49]. The WD masses in the Milky Way from the SDSS data show nearly a Gaussian distribution, peaked at 0.6 $M_\odot$ [50]. Assuming that the mass distribution of WD in NS-WD binaries is similar to the SDSS result and that the massive WD (> 0.6 $M_\odot$)-NS (1.4 $M_\odot$) binaries (total mass > 2.0 $M_\odot$) will mostly form magnetars, the magnetar formation rate will be roughly < 50% of the NS-WD merger rate. Here we omit the very massive WDs ($\geq 1.1\ M_\odot$, the binaries $\geq 2.5\ M_\odot$ which will form BHs) and small mass WDs $\leq 0.2\ M_\odot$, because both of them take only a very small/negligible fraction in the WD mass distribution.

Therefore, the magnetar formation rate in the channel of the NS-WD merger is about 20 times less than type Ia SN rate and about 10 times less than the classical core-collapse magnetar formation rate of $\sim 3 \times 10^4$ $Gpc^{-3}yr^{-1}$ in $z < 0.5$ [51], but larger than the magnetar formation rate from the binary NS mergers which is $1540^{+3200}_{-1200}\ Gpc^{-3}yr^{-1}$ [52]. The magnetar formation rate in the channel of the AIC of WD or binary WDs is highly uncertain probably comparable to that of the binary NS mergers [53].

If the magnetar can produce repeating FRBs at a very young age, we will expect that the birth rate of the repeating FRB source should be proportional to the magnetar formation rate. Likewise, repeating FRBs will hold different properties (mass, spin, $B$-field, and magnetosphere, nebula, etc.) if their magnetars originate from the different channels.

The magnetar formed in the channel of massive WD-NS mergers in our model should consist of even more `dirty' (or denser) magnetosphere/nebula, which contains enormous degenerate electrons inflated from the WD during the collapse of the merger. The dense plasma can lead to an extreme and dynamic magne-to-ionic environment. The high and variable Faraday rotation measure of $\sim 10^5$ rad/s observed in the repeating FRB 121102 [19] is indeed supportive of our model.

## 4. Summary

There are three different remnants of the NS-WD mergers. The first kind is a stellar mass BH. The second kind is a new NS which can be formed from the NS-WD merger product when the total mass is sufficiently large. We propose that some of the new NSs may form magnetars, assuming that the core of the WD collapses into the NS and an enormous quantity of the degenerate electrons of the WD collapses into the outer layers of the new NS. That will lead to the electrons and positive charges distributed differently with different angular velocities by the differential rotation of the newborn NS, and in this way, strong magnetic fields can be formed for a magnetar.

Repeating FRBs can be produced by the electrons energized by the magnetic reconnection in the newborn magnetar, due to the movements of the magnetar's crustal plate or starquakes. The magnetar formed in the channel of massive WD-NS mergers is more `dirty' with denser magnetosphere/nebula which contains enormous degenerate electrons inflated from the WD during the final collapse of the merger. These dense plasmas will be responsible for the extreme and dynamic magneto-ionic environment as observed in the first repeating FRB 121102.

The magnetar formation rate in the new channel is roughly estimated to be 10 times less than the classical core-collapse magnetar formation rate, but it would be larger than the magnetar formation rate from the binary NS mergers.

The NS-WD merger product will not collapse to form a BH or a new NS if the binary mass is not large enough. Then the remaining hybrid WDNS could be a kind of new long-lived object. Multi-wavelength emissions will be expected from such an object.

## Acknowledgements

This research was funded by the National Key R&D Program of China, under grant number 2018YFA0404602. I thank the anonymous referees and Binbin Zhang for helpful comments, which improved the paper significantly.

## Conflicts of Interest

The author declares no conflicts of interest regarding the publication of this paper.